\documentclass[11pt]{amsart}
\usepackage{latexsym,amssymb,amsmath,amscd,amsthm}
\usepackage{amsfonts}
\numberwithin{equation}{section}
\theoremstyle{remark}

\newcommand{\bq}{\begin{equation}}
\newcommand{\bea}{\begin{array}}
\newcommand{\eea}{\end{array}}

\newcommand{\ga}{\alpha}
\newcommand{\gep}{\epsilon}
\newcommand{\gD}{\Delta}
\newcommand{\gl}{\lambda}
\newcommand{\gL}{\Lambda}
\newcommand{\gb}{\beta}

\newcommand{\mf}{\mathfrak}
\newcommand{\mc}{\mathcal}

\newcommand{\go}{\omega}
\newcommand{\gO}{\Omega}
\newcommand{\gG}{\Gamma}

\newcommand{\gs}{\sigma}

\newcommand{\gag}{\gamma}
\newcommand{\gd}{\delta}
\newcommand{\pp}{\partial}

\newcommand{\tl}{\tilde}
\newcommand{\na}{\nabla}
\newcommand{\gk}{\kappa}

\newcommand{\bs}{\blacksquare}

\newcommand{\gS}{\Sigma}

\newcommand{{\DDD}}{D\!\!\!\!\!\!-}

\newcommand{\bx}{\Box}

\title{REMARKS ON OSMOSIS, QUANTUM MECHANICS, AND
GRAVITY}

\author{Robert Carroll\\University of Illinois, Urbana, IL 61801}

\date{April, 2011\thanks{email: rcarroll@math.uiuc.edu}}

\begin{document}


\bibliographystyle{plain}


\begin{abstract} 
Some relations of the quantum potential to Weyl geometry
are indicated with applications to the Friedmann equations for
a toy quantum cosmology.  Osmotic velocity and pressure are
briefly discussed in terms of quantum mechanics and superfluids
with connections to gravity.
\end{abstract}

\maketitle


\section{REMARKS ON WEYL GEOMETRY}
\renewcommand{\theequation}{1.\arabic{equation}}
\setcounter{equation}{0}

We begin with some features of Weyl geometry and the Dirac-Weyl
theory following \cite{isrt,istr,isrn,rosn} (cf. also \cite{adsr,audr,agsn,aula,bqcd,
caht,c006,c007,c009,caro,cl,crar,clrr,cstr,drlr,papn,quir,sant,sclz,shoj,whlr}).  Note that the 
paper is in part an expansion of some ideas in \cite{clrr}
and we correct some notational confusion from \cite{c009,clrr}.
There is a curious issue here of too much or too little, which is compounded by various, sometimes conflicting, notations and we feel that it is
well advised to simply follow the extensive development of Israelit-Rosen which will keep the notation under control
and benefit from insights therein.
The background involves looking at a metric tensor $g_{ab}=
g_{ba}$ and a length connection vector $w_{\mu}$.  If a vector
is displaced by $dx^{\nu}$ then
\bq\label{1.1}
dB^{\mu}=-B^{\gs}\hat{\gG}^{\mu}_{\gs\nu}dx^{\nu};\,\,
dB=Bw_{\nu}dx^{\nu}
\end{equation}
One can also write $({\bf 1A})\,\,d(B^2)=2B^2w_{\nu}dx^{\nu}$
and this all requires that $({\bf 1B})\,\,g_{\mu\gs}\hat{\gG}^{\gs}_{\nu\gs}+g_{\nu\gs}\hat{\gG}^{\gs}_{\mu\gs}=\pp_{\gs}g_{\mu\nu}-2g_{\mu\nu}w_{\gs}$.  Then assuming that $\hat{\gG}^{\gs}_
{\mu\nu}=\gG^{\gs}_{\mu\nu}$ there results
\bq\label{1.2}
\hat{\gG}^{\gs}_{\mu\nu}=\gG^{\gs}_{\mu\nu}+g_{\mu\nu}w^{\gs}-\gd^{\gs}_{\nu}w_{\mu}-\gd^{\gs}_{\mu}w_{\nu}
\end{equation}
where $\gG^{\gs}_{\mu\nu}$ denotes the standard Riemannian
Christoffel symbol.  The covariant derivative for $g_{ab}$ is written $\na_{\ga}$ or e.g. $B^{\mu}_{;\nu}=\na_{\nu}B^{\mu}$
(and involves $\gG^{\gs}_{\mu\nu}$) whereas
with the Weyl connection (1.2) one forms
a co-covariant derivative of a vector
\bq\label{1.3}
B_{:\nu}^{\mu}=\pp_{\nu}B^{\mu}+B^{\gs}\hat{\gG}^
{\mu}_{\gs\nu};\,\,B_{\mu,:\nu}=\pp_{\nu}B_{\mu}-B_{\gs}\hat{\gG}^{\gs}_{\mu\nu}
\end{equation}
so that
\bq\label{1.4}
B^{\mu}_{:\nu}=B^{\mu}_{;\nu}+B^{\mu}(g_{\gs\nu}w^{\mu}-\gd^{\mu}_{\nu}w_{\gs}-\gd^{\mu}_{\gs}w_{\nu})
\end{equation}
Note also
\bq\label{1.5}
g_{\mu\nu:\gs}=2g_{\mu\nu}w_{\gs};\,\,g^{\mu\nu}_{:\gs}=-2g^{\mu\nu}w_{\gs}
\end{equation}
Now under Weyl gauge transformations one has a length change
$({\bf 1C})\,\,B\to\hat{B}\sim\tl{B}=e^{\gl(x)}B$ with $g_{\mu\nu}\to
\tl{g}_{\mu\nu}=e^{2\gl}g_{ab}$ and $g^{ab}\to e^{-2\gl}g^{ab}$.
Writing $e^{2\gl(x)}=f(x)$ there is a calculation from \cite{dirc}
for lengths $\ell\to\hat{\ell}=f^2\ell$
\bq\label{1.6}
2\hat{\ell}d\hat{\ell}\sim f_{\ga}dx^{\ga}\ell^2+2f\ell d\ell\Rightarrow
\frac{d\hat{\ell}}{\hat{\ell}}\sim\frac{1}{2}\pp_{\ga}log(f)dx^{\ga}+
w_{\ga}dx^{\ga}\Rightarrow\hat{w}_{\ga}=
\end{equation}
$$=w_{\ga}+\frac{1}{2}\pp_{\ga}log(f)=w_{\ga}+\pp_{\ga}\gl(x)$$

\indent
{\bf REMARK 1.1.}
In Weyl geometry consider now a scalar S of Weyl weight n defined via
$\tl{S}=exp(n\gl)S$ (e.g. $B=(g_{ab}B^aB^b)^{1/2}=\ell\to\hat{\ell}=e^{\gl}\ell$) and $\pp_{\mu}\tl{S}=exp(n\gl)(\pp_{\mu}S+nS\pp_{\mu}\gl)$.  This is not covariant with respect to
Weyl gauge transformations (WGT) unless $n=0$ but one can define a gauge covariant derivative $({\bf 1D})\,\,S_{||\mu}=\pp_{\mu}S-nSw_{\mu}$ with $\tl{S}_{||\mu}=exp(n\gl)S_{||\mu}$
(using (1.6)).
For a gauge covariant derivative of a vector $S^{\mu}$ of Weyl weight n, where $\tl{S}^{\ga}=
exp(n\gl)S^{\ga}$, there results $({\bf 1E})\,\,S^{\mu}_{||\nu}=S^{\mu}_{:\mu}-nS^{\mu}w_{\nu}=exp(n\gl)S^{\mu}_{||\nu}$
(note $\hat{\gG}^{\mu}_{\gs\nu}-n\gd^{\mu}_{\gs}w_{\nu}$ is actually a connection).  $\bs$
\\[3mm]\indent
Dirac \cite{dirc} (1973) introduced a $\gb$ field of weight $-1$
(as a Lagrange type multiplier) in order to have a satisfactory
scalar tensor action of the form (cf. \cite{isrt}, Chap. 4)
\bq\label{1.7}
I=\int\sqrt{-g}d^4x[-\gb^2R+\gs\gb^2w^cw_c+(\gs+6))\pp_c\gb\pp^c\gb+2\gs\gb\pp_c\gb w^c+2\gL\gb^4]
\end{equation}
Note here that $\sqrt{-g}$ is of weight 4 and R of weight -2;
electromagnetic terms have been omitted (also $\hbar=c=1$).
It is this factor
$\gb$ which was identified as a quantum mass ${\mf M}=m
exp({\mf Q})$ in \cite{shoj} (and subsequently utilized in \cite{c006,
c007,c009,caro,cl,crar,clrr}).  It's quantum significance was also
noticed in \cite{bqcd} (cf. also \cite{quir}) in connection
with conformal GR (GR stands for general relativity).  
Unfortunately a different notation was used in \cite{bqcd} and in
\cite{quir} (0009169), namely $({\bf 1F})\,\,\hat{g}_{ab}=\gO^2(x)g_{ab}$ with $\gO^2=exp(-\psi)$ in \cite{bqcd} and $\gO^2=
exp(\psi)$ in \cite{quir} (0009109) - a fact which I occasionally
overlooked in subsequent work (cf. \cite{clrr} for some errata).  
The latter paper \cite{quir}
(0009169) contains some important material relating various
actions in Einstein frame (EF) and Jordan frame (BD type
equations).  It was shown that an action 
\bq\label{1.8}
S_4=\int d^4x\sqrt{-\hat{g}}e^{-\psi}\left[\hat{R}-\left(\ga-
\frac{3}{2}\right)
|\hat{\na}\psi|^2+16\pi e^{-\psi}L_M\right]
\end{equation}
was the only one (of 4 considered) which was invariant under
transformations of units (length, time, and mass) and this was
then dubbed a string frame action (under the assumption that
strings are ultimate and inevitable - cf. \cite{grne}).  Note that $\gO^2=exp(\psi)
=\phi$ implies
\bq\label{1.9}
\hat{g}_{ab;c}=\pp_c\psi\hat{g}_{ab}\Rightarrow w_c=\frac
{1}{2}\pp_c\psi
\end{equation}
In particular conformal Riemannian geometry as in $S_4$ is a
WIST (Weyl integrable space time) as noted in \cite{quir} 
(0009169) and \cite{novl,nsfo}.
\\[3mm]\indent
Now $S_4$ arises in our work via a conformal map 
$g_{ab}\to\hat{g}_{ab}=\gO^2g_{ab}$ from an EF action
(cf. \cite{mgsk}) $({\bf 1G})\,\,S=\int \sqrt{-g}d^4x[R-\ga
|\na\psi|^2+16\pi L_M]$ in the form $({\bf 1H})\,\,\hat{S}=\int
\sqrt{-\hat{g}}d^4x[\hat{\phi}\hat{R}-(\go/\hat{\phi})|\hat{\na}\hat{\phi}|^2+16\pi\hat{\phi}^2L_M]$ where we will write out the theory here in terms of $\gO^2=\exp(\psi)$ following \cite{caro}
(cf. also \cite{c006,c007} and note a few sign adjustments are
needed in \cite{c009}).  Recall first that $\phi\sim\gb^2/m^2=
{\mf M}^2/m^2$ and $\hat{\phi}=\phi^{-1}=exp(-\psi)$ and note that
$\sqrt{-g}=\hat{\phi}^2\sqrt{-\hat{g}}$ (cf. \cite{c006,c007}) and
\bq\label{1.10}
\sqrt{-\hat{g}}\hat{\phi}\hat{R}=\hat{\phi}^{-1}\sqrt{-\hat{g}}\hat{\phi}^2\hat{R}=\hat{\phi}^{-1}\sqrt{-g}\hat{R}=\frac{\gb^2}{m^2}
\sqrt{-g}\hat{R}
\end{equation}
But via \cite{dirc,isrt} we can write (cf. also \cite{c009} for the pattern here)
\bq\label{1.11}
\gb^2\hat{R}=\gb^2R-6\gb^2\na_{\gk}w^{\gk}+6\gb^2w^{\gk}w_{\gk}
\end{equation}
and via$-\gb^2\na_{gk}w^{\gk}=-\na_{\gk}(\gb^2w^{\gk})+
2\gb\pp_{\gk}\gb w^{\gk}$, with a vanishing divergence term
upon integration, the first integral in $\hat{S}$ becomes
\bq\label{1.12}
I_1=\int\sqrt{-g}d^4x\left[\frac{\gb^2}{m^2}R+12\gb\pp_{\gk}\gb w^{\gk}+6\gb^2w^{\gk}w_{\gk}\right]
\end{equation}
The second integral is ($\gag=\ga-(3/2)$)
\bq\label{1.13}
I_2=-\gag\int \sqrt{-\hat{g}}d^4x\hat{\phi}\frac{|\hat{\na}\hat{\phi}|^2}{|\hat{\phi}|^2}=-\frac{4\gag}{m^2}\int\sqrt{-g}d^4x|\hat{\na}\gb|^2
\end{equation}
since $\hat{\phi}_c=-\psi_ce^{-\psi},\,\,\hat{\na}\psi=-\hat{\na}
\hat{\phi}/\hat{\phi},\,\,\hat{\phi}=\phi^{-1}=m^2\gb^{-2},$ and 
$\hat{\phi}/\hat{\phi}=-2\hat{\na}\gb/\gb$ (along with (1.10)).
Via ({\bf 1D}) $\hat{\na}_c\gb=\pp_c\gb+w_c\gb$ for $\Pi(\gb)=-1$ ($\Pi$ denotes the Weyl weight) and from $\gb^2=m^2exp(\psi)$ one has $\psi_cexp(\psi)=(\gb^2/m^2)\psi_c\Rightarrow
\psi_c=2\gb_c/\gb$.  Hence via (1.9) $w_c=\gb_c/\gb$ 
(correcting \cite{c009}) and consequently $\hat{\na}_c\gb=
2\gb_c$.  In any event
\bq\label{1.14}
\hat{S}_{GR}=\frac{1}{m^2}\int\sqrt{-g}d^4x[\gb^2R+12\gb\pp_c
\gb w^c+6\gb^2w^cw_c-4\gag|\hat{\na}\gb|^2+16\pi m^2L_M]
\end{equation}
(as in \cite{c009}, p. 59 (6.6) and p. 237 (3.32)).
\\[3mm]\indent
{\bf REMARK 1.2.}
There was some sign confusion in \cite{c009} which we have adjusted above.  Thus equations (6.7)-(6.8) on p. 59 and
(3.33))-(3.34) on p. 237 should be altered as indicated below.
$\bs$
\\[3mm]\indent
First we write $|\hat{\na}\gb|^2 = \gb^c\gb_c$ and from (1.9)
and (1.14)
\bq\label{1.15}
I_1=\frac{1}{m^2}\int\sqrt{-g}d^4x[\gb^2R+18\pp_c\gb
\pp^c\gb];\,\,I_2=-16\gag
\int \sqrt{-g}d^4x\gb^c\gb_c
\end{equation}
Consequently (recall $\gag=\ga-(3/2)$)
\bq\label{1.16}
\hat{S}=\frac{1}{m^2}\int \sqrt{-g}d^4x[\gb^2R+16\pi m^2L_M+
2(21-8\ga)\gb^c\gb_c]
\end{equation}
Then we rewrite (1.7) in the form
\bq\label{1.17}
I=\int \sqrt{-g}d^4x[-\gb^2R+3(\gs+2)\gb_c\gb^c+2\gL\gb^4]
\end{equation}
Ignoring $L_M$ and $\gL$ for the moment we see that 
$I=m^2\hat{S}$ provided that $({\bf 1I})\,\,3\gs=8(3-\ga)$ and,
a priori, $\ga$ or $\gs$ can be arbitrary (modulo e.g. possible cosmological restrictions on $\ga$).
\\[3mm]\indent
{\bf REMARK 1.3}
The Brans-Dicke (BD) type theory of $S_4$ (written also as $\hat{S}_4$)
was used for the Friedman equations in \cite{clrr} for example
where a relation $\gs\sim -4\ga$ was used (based on some
sign confusion).  
Hence one
should convert the Weyl-Dirac action (1.1) of \cite{clrr} to (1.7)
and use the $(\gs,\ga)$ relation ({\bf 1I}) in order to obtain
the correct Friedmann equations for the toy quantum
cosmology considered there (cf. Section 2).
$\bs$

\section{REMARKS ON COSMOLOGY}
\renewcommand{\theequation}{2.\arabic{equation}}
\setcounter{equation}{0}

In \cite{clrr} the Friedmann equations were based on an $S_4$
type action
\bq\label{2.1}
\tl{S}=\frac{1}{16\pi}\int \sqrt{-g}d^4x\left[R\Phi-\go\frac{|\na\Phi|^2}{\Phi}+L_M\right]
\end{equation}
where $({\bf 2A})\,\,L_M=-V(\Phi)+16\pi{\mf L}$ with $\gL$ assumed to be suitably inserted in ${\mf L}_M$ (cf. \cite{faro,
fumd}).  An FRW metric ($c=1$)
\bq\label{2.2}
ds^2=dt^2-a^2(t)\left[\frac{dr^2}{1-kr^2}+r^2d\gS^2\right]
\end{equation}
is used and one has field equations as in \cite{clrr}
\bq\label{2.3}
G_{ab}=\frac{8\pi}{\Phi}T^M_{ab}+\frac{\go}{\Phi^2}\left[\na_a\Phi\na_b\Phi -\frac{1}{2}
g_{ab}\na^c\Phi\na_c\Phi+\right]
\end{equation}
$$+\frac{1}{\Phi}(\na_a\na_b\phi-g_{ab}\bx\Phi)-\frac{V}{2\Phi}g_{ab}$$
where $({\bf 2B})\,\,T^M_{ab}=-(2/\sqrt{-g})(\gd/\gd g^{ab})(\sqrt{-g}{\mf L}_M)$ (we ignore a possible shift $\sqrt{-g}\to\sqrt{g}$,
etc. due to the signature in (2.2)). 
Variation of the action with
respect to $\Phi$ gives
\bq\label{2.4}
\frac{2\go}{\Phi}\bx\Phi+R-\frac{\go}{\Phi^2}\na^c\Phi\na_c\Phi-\frac{dV}{d\Phi}=0
\end{equation}
with trace
\bq\label{2.5}
R=-\frac{8\pi T_M}{\Phi}+\frac{\go}{\Phi^2}\na^c\Phi\na_c\Phi+\frac{3\bx\Phi}{\Phi}+\frac{2V}{\Phi}
\end{equation}
and using (2.5) to eliminate R from (2.4) yields
\bq\label{2.6}
\bx\Phi=\frac{1}{2\go+3}\left[8\pi T^M+\Phi\frac{dV}{d\Phi}-2V\right]
\end{equation}
Then using (2.2) one obtains $({\bf 2C})\,\,\na^c\Phi\na_c\Phi=-(\dot{\Phi})^2$ and $\bx\Phi=-(\ddot{\Phi}+3H\dot{\Phi})=-(1/a^3)(d/dt)(a^3\dot{\Phi})$.  Assume now that $({\bf 2D})\,\,T^M_{ab}=(P^M+\rho^M)u_au_b+P^Mg_{ab}$ and then the time dependent component of the BD field equations
gives a constraint equation
\bq\label{2.7}
H^2=\frac{8\pi}{3\Phi}+\frac{\go}{6}\left(\frac{\dot{\Phi}}{\Phi}\right)^2-H\frac{\dot{\Phi}}{\Phi}-
\frac{k}{a^2}+\frac{V}{6\Phi}
\end{equation}
Then, using $R=6[\dot{H}+2H^2+(k/a^2)]$, there results
\bq\label{2.8}
\dot{H}+2H^2+\frac{k}{a^2}=-\frac{4\pi T^M}{3\Phi}-\frac{\go}{6}\left(\frac{\dot{\Phi}}{\Phi}\right)^2
+\frac{1}{2}\frac{\bx\Phi}{\Phi}+\frac{V}{3\Phi}
\end{equation}
From (2.6), (2.7), and the trace equation $({\bf 2E})\,\,T^M=3P^M-\rho^M$ one has then
\bq\label{2.9}
\dot{H}=\frac{-8\pi}{(2\go+3)\Phi}\left[(\go+2)\rho^M+\go P^M\right]-\frac{\go}{2}\left(\frac
{\dot{\Phi}}{\Phi}\right)^2+
\end{equation}
$$+2H\frac{\dot{\Phi}}{\Phi}+\frac{k}{a^2}+\frac{2}{2(2\go+3)\Phi}\left(\Phi\frac{dV}{d\Phi}
-2V\right)$$
and (2.6) reduces to
\bq\label{2.10}
\ddot{\Phi}+3H\dot{\Phi}=\frac{1}{2\go+3}\left[8\pi(\rho^M-3P^M)-\Phi\frac{dV}{d\Phi}+2V\right]
\end{equation}
\\[3mm]\indent
In order to apply this to our model (1.7) we look at these Friedmann equations for $\Phi=\hat{\phi}$ (with ${\mf Q}\sim\psi$)
where $(\gb^2/m^2)=\hat{\phi}^{-1}=exp(\psi)=exp({\mf Q})$.  Then $\hat{\phi}=exp(-{\mf Q})=\Phi$  and equations (2.3)-(2.10) can be written in terms of the quantum potential  ${\mf Q}$ from $({\bf 2F})\,\,{\mf Q}=(\hbar^2/m^2)(\bx|\Psi|/|\Psi|)$ with $\Psi$
arising in the quantum process underlying (1.2) (cf. \cite{audr,agsn,aula,c006,c007,c009,cl,crar,clrr,cstr,sant,shoj,whlr}
and see also \cite{brfm,fgme}).
We assume first that ${\mf L}_M=0$ and then one can write $({\bf 2G})\,\,\dot{\Phi}=-\dot{{\mf Q}}
\Phi,\,\,\ddot{\Phi}=(\dot{{\mf Q}}^2-\ddot{{\mf Q}})\Phi$ and thence e.g.
\bq\label{2.11}
(\dot{{\mf Q}}^2-\ddot{{\mf Q}})\Phi-3H\dot{{\mf Q}}\Phi=\frac{1}{2\go+3}\left[8\pi
(\rho^M-3P^M)-\Phi\frac{dV}{d\Phi}+2V\right]
\end{equation}
Note that $({\bf 2H})\,\,\go=\ga-(3/2)$ and $3\gs=8(3-\ga)\Rightarrow \gs=12-(8/3)\go$.  
We see that the dynamics of $\Phi$ is determined in part by $V(dV/d\Phi)-2V$ (which vanishes
for $V=c\Phi^2=cexp(-2{\mf Q})$) and by $\rho^M,\,\,P^M$.  Thus, in
particular, it is possible to envision some cosmological behavior provided by a quantum background as in (1.7).  We note also that Mannheim refers to intrinsically quantum mechanical gravity associated to general Weyl geometry with the Weyl tensor, etc. (cf. \cite{bdmh,mnhm}).  We especially
recommend Padmanabhan's paper \cite{padn} (1012.4476) 
and references there for
a discussion of quantum mechanics and spacetime.
Even more striking is the discussion in Gr\"ossing's book \cite{grsg} (p. 123) which indicates (assuming an aether) how Einsteinian gravity can be considered a pure quantum phenomenon.
\\[3mm]\indent
{\bf REMARK 2.1.}
It is interesting to note that via $\hat{\phi}=exp(-{\mf Q})$
one has ${\mf Q}\sim\psi$ and hence by (1.9) the Weyl vector
is $w_c=(1/2)\pp_c{\mf Q}$ inplying a geometric role for 
${\mf Q}$.  In particular this seems to show a direct connection
of quantum mechanics with gravity.  Of course this is not 
surprising since the geometry of (1.7) was generated via 
quantum mechanics but it suggests that Weyl geometry itself
may have an implicit quantum connection via $w_a=(1/2)\pp_a
{\mf Q}$ with ${\mf Q}$ a quantum potential.  This seems to be a 
new observation (although connections of QM to the Weyl-
Ricci curvature arise (cf. \cite{audr,agsn,c006,c007,cl,cstr,sant,whlr}).  $\bs$
\\[3mm]\indent
{\bf REMARK 2.2.}
In \cite{shoj} there are also joint field equations involving the conformal factor $\gO^2$ (expressed via $\phi$ or $\phi^{-1}$) and for the quantum potential ${\mf Q}$ (expressed separately via a Lagrangian) with the goal of thereby deriving connections between ${\mf Q}$ and $\phi$.
Also Bohmian aspects of the theory are emphasized.  Some of this is also reviewed in \cite{c006,
c007}.  A main conclusion of this is that the quantum potential is a dynamical field and interactions
between $\gL$ and ${\mc Q}$ represent a connection between large and small scale 
structures.  We note also from 
\cite{novl} that $\gL$ plays an important role in generating mass for scalar fields and the understanding of $\gL$ is considerably
deepened and enhanced via work of Klinkhamer and Volovik
(cf. \cite{klhr,klvk,volv,vovk}) some of which is described briefly
below.
$\bs$
\\[3mm]\indent
{\bf REMARK 2.3.}
Relations between thermodynamics and gravity have been extensively studied following early
work of Bekenstein and Hawking plus more recently Jacobson \cite{jacb}, and Padmanabhan
\cite{pada,padn}.  One can derive general Einstein field equations via thermodynamic
principles and gravity itself seems to be characterized via thermodynamics (see \cite{c009}
for a brief sketch of some of this and there is much more information in works of Padmanabhan
et al.  The work of Verlinde \cite{vlde} has triggered another explosion of interest in entropy
and gravity and we mention here only a few articles,
e.g. \cite{cai,caoh,ccsh,wlwg} (related to the Friedmann equations) and \cite{makl,mkmr,nels,smol} (relations to quantum mechanics).  $\bs$ 
\\[3mm]\indent
We would like to mention here that entropy in quantum mechanics is normally connected
to momentum fluctuations and the quantum potential gives rise to an entropy functional
$\int P{\mf Q}\,dx$ corresponding to Fisher information (here $P=|\psi|^2$ where $\psi$ is a wave function and ${\mf Q}$ represents a 3-dimensional quantum potential).  Generally
${\mf Q}$ can be described e.g. via an osmotic velocity or a thermalization of this (cf. \cite{c006,c007,c009,crow,fred,garb,gros,hall,hlrg,hkrg} and Section 3).  There is also a gravitational version of this
related to the Wheeler-deWitt (WDW) framework in the form
\bq\label{2.12}
\int {\mc D}h\,P{\mf Q}=\int {\mc D}h\frac{\gd P^{1/2}}{\gd h_{ij}}G_{ijk\ell}\frac{\gd P^{1/2}}
{\gd h_{k\ell}}
\end{equation}
(cf. \cite{c007,c009,crcl,fred,hkrg}) and see also \cite{glkn,wanc}
for WDW).    We note also the Perelman entropy functional
\bq\label{2.13}
{\mf F}=\int_M(R+|\na f|^2)e^{-f}dV
\end{equation}
and corresponding Ricci flows are related to a so called Nash entropy $({\bf 2I})\,\,S=
\int u\,log(u)\,dV$ where $u=exp(-f)$ and various aspects of quantum mechanics related to
the Schr\"odinger equation and Weyl geometry arise (cf. \cite{c009,caar,caro,graf,isfc,khol}).

\section{SOME BACKGROUND}
\renewcommand{\theequation}{3.\arabic{equation}}
\setcounter{equation}{0}

First it is interesting and perhaps enlightening to have some 3-D
background here on the quantum potential.  Thus, referring
to \cite{c006,c007,c009,crrl,crar,cath,fred,fpps,frgy,fnsf,garb,
gros,gfps,gsps,hall,hlrg,hkrg,naga,nels,regn,relg,smol}
we recall some results involving the QP relative to a 3-D SE.
Various notations are used and we deal with a Schr\"odinger
equation (SE) $({\bf 3A})\,\,i\hbar\psi_t=\gD\psi+V\psi$ with 
$\psi=Rexp(iS/\hbar)$ and 
\bq\label{3.1}
\pp_tS+\frac{1}{2m}|\na S|^2+V+{\mf Q}=0;\,\,\pp_t(R^2)+\frac{1}{m}\na(R^2\na S)=0
\end{equation}
where the quantum potential (QP) is defined via
\bq\label{3.2}
{\mf Q}=-\frac{\hbar^2}{2m}\frac{\gD R}{R}=-\frac{\hbar^2}{4m}
\left[\frac{1}{2}\left(\frac{\na P}{P}\right)^2-\frac{\gD P}{P}\right]
\end{equation}
(note this is different from ${\mf Q}$ in e.g. ({\bf 2F})).
Here $|\psi|^2=R^2=P$ and $\rho=mP$.  Setting $D=(\hbar/2m)$
one can write $({\bf 3B})\,\,{\bf u}=-D(\na P/P)$ (osmotic velocity)
which is expressed in terms of a ``canonical" momentum fluctuation $\gd p/m={\bf u}$ (cf. \cite{c006,c007,c009} for references and more details).  Now setting ${\bf k}_u=-(1/2)
{\bf u}$ one can write $({\bf 3C})\,\,{\mf Q}=(\hbar^2/2m)(|{\bf k}_u|^2-\na\cdot{\bf k}_u)=(m/2){\bf u}^2-(\hbar/2)(\na\cdot{\bf u})$.
There is also a thermalization of the QP based on \cite{gros}
(cf. also \cite{c009,crrl}) whereby one refers to a basic particle
energy $E=\hbar\go$ (where $kT=\hbar\go=1/\ga$ with $k\sim
k_B$ the Boltzman constant).
One then defines a thermal function ${\mc Q}$
where $\gD{\mc Q}={\mc Q}(t)-{\mc Q}(0)<0$ determines a heat
dissipation with an osmotic velocity $({\bf 3D})\,\,{\bf u}=-
D(\na P/P)=(1/2\go m)\na{\mc Q}$.  Hence and from (3.2), ({\bf 3B}), and ({\bf 3D}) there results
\bq\label{3.3}
\int {\mf Q}Pd^3x=\frac{\hbar^2}{8m}\int\left(\frac{\na P}{P}\right)^2Pd^3x
=\frac{m}{2}\int |{\bf u}|^2Pd^3x=
\end{equation}
$$=\frac{1}{8\go^2m}\int
|\na{\mc Q}|^2Pd^3x=\tl{c}(t)\int (\na{\mc Q})^2exp^{-\ga{\mc Q}}d^3x$$
(the last equation via $log(P)=-\ga{\mc Q}+c(t)\Rightarrow
P=\hat{c}(t)exp(-\ga{\mc Q})$ where $\ga=1/\hbar\go$ -
cf. \cite{c009,crrl,gros,gfps,gsps} for more on this).  One
should also note that $({\bf 3E})\,\,\int P{\mf Q}d^3x=
(\hbar^2/8m)\int[(|\na P|^2/P]d^3x=(\hbar^2/8m)FI$ where FI
denotes Fisher information.
\\[3mm]\indent
Now we note that an idea of osmotic pressure arises in the
Klinkhamer-Volovik theory of superfluids (cf. \cite{klhr,klvk,volv,
vovk}) from which we extract here (see especially gr-qc 0711.3170, 0806.2805, 0907.4887, 0811.4347, and 1102.3152;
hep-th 0907.4887 and 1101.1281; cond-mat 1004.0597;   
physics 0909.1044 and cf. also \cite{ezaw,hark,ptsh,pist}).
This is a relativistic theory and the relativistic quantum
vacuum is considered as a self sustained medium described by a variable ${\mf q}$ which is a conserved quantity analogous to a particle density $n$ in condensed matter theory but ${\mf q}$ is the
relativistic invariant quantity (see also Section 4).  There is a Gibbs-Duhem relation
$\gep_{vac}({\mf q})-\mu {\mf q}=-P_{vac}$ where $\gep_{vac}$ is the 
vacuum energy density and $\mu$ is the vacuum chemical
potential (thermodynamically conjugate to ${\mf q}$).  Dynamical
equations for ${\mf q}$ show that $\gL=\gep_{vac}-\mu {\mf q}=-P_{vac}$ 
so the vacuum variable is automatically self-tuned to nullify
in equilibrium any contribution from different quantum fields.
In \cite{vovk} (0909.1044) one considers a dilute solution of 
${}^3He$ quasiparticles in a superfluid vacuum ${}^4He$ at
$T=0$.  The negative contribution of the vacuum to the osmotic
pressure of ${}^3He$ is given by the same equation 
\bq\label{3.4}
P_{os}=P_{mat}+P_{vac}=P_{mat}-\frac{1}{2}\chi_{vac}\left[
{\mf q}\frac{\pp\gep_{mat}(n,{\mf q})}{\pp {\mf q}}\right]^2
\end{equation}
where $\chi_{vac}$ is the vacuum compressibility
\bq\label{3.5}
\chi_{vac}^{-1}=\left[{\mf q}^2\frac{d^2\gep_{vac}
({\mf q})}{d{\mf q}^2}\right]_{{\mf q}={\mf q}_0}\geq 0
\end{equation}
(where now $\chi_{vac}$ corresponds to the compressibility
of liquid ${}^4He$.  This negative contribution to the osmotic
pressure has been experimentally verified (according to 
0909.1044).
\\[3mm]\indent
{\bf REMARK 3.1 .}
Note that the formula in (3.3) involving 
the 3-D integral
$\int |\na{\mc Q}|^2exp(-\ga{\mc Q})d^3x$ reminds one of the 
$I_2$ formula (1.13) from $S_4$ where there is a 4-D integral
involving $exp(-\psi)|\hat{\na}\psi|^2$; also the $S_4$ or
$I_2$ represent ``action" terms in the Weyl theory as does
$\int P{\mf Q}d^3x$ in the 3-D theory (cf. \cite{c006,c007,c009}).
However there is no immediate extrapolation here from 
3-D to 4-D or
vice-versa.  In fact we know from remarks before (2.11) that
$\psi\sim {\mf Q}$ in the 4-D theory where ${\mf Q}$ is given
by ({\bf 2F}) 
and in the 3-D theory there
is a formula (cf. \cite{c009,gros})
\bq\label{3.6}
{\mf Q}=-\frac{\hbar^2}{4k_BTm}\left[\na^2{\mc Q}-\frac{1}{D}\pp_t{\mc Q}\right]
\end{equation} 
We refer to Section 4 for some remarks on the KG equation
and recall that there is related information via the Wheeler-deWitt
(WDW) or Klein-Gordon (KG) equations in e.g. \cite{brfm,c009,
crcl,fgme,fred,grsg,hkrg}.
 $\bs$
\\[3mm]\indent
{\bf REMARK 3.2.}
The role of information theory and entropy in QM has been
clarified by e.g. Caticha \cite{cath} (see also in this connection
\cite{bmhl,naga,nels,smol}).  Osmotic velocity plays an important
role in the diffusion theory of course and we mention also some
recent work of Munkhammar \cite{mkmr} related to \cite{vlde}.
One can argue that sometimes QM is signaled simply by the 
entrance of $\hbar$ into the theory.  For example in the holographic force scenario of Verlinde \cite{vlde} it seems to
arise via the Unruh temperature relation to acceleration
$({\bf 3F})\,\,k_BT=(1/2\pi)(\hbar a/c)$.  Padmanabhan et al
have written deeply and extensively about matters of entropy,
entanglement, equipartition, horizons, and gravity as an
emergent phenomenon (cf. \cite{pada,padn}).  Entanglement entropy is also studed by J.W. Lee et al in \cite
{klle,leew,lkle} and one theme involves entropy and information
erasure.  An interesting Machian point of view is developed in
\cite{ggbv,gbkv} and in \cite{xiog} there is a microscopic
approach to entropic gravity via the idea of coherence 
length.  In \cite{gros} one defines a thermal term $s$ via
$\na{\mc Q}=2\go\na(\gd s)=2\go m{\bf u}=-\go\hbar(\na P/P)$,  
providing the osmotic velocity, and in \cite{cath} this is combined with a relative entropy term ${\mf S}$,
based on ``hidden" variables, for the ``drift" velocity
and a current velocity determined via $\phi$ where $\phi={\mf S}
-log(\rho^{1/2})$ with velocity connection $v^a=b^a+u^a$
(see \cite{cath} for details).  In \cite{joct} it is also shown that in 
entropic quantum dynamics gravitational potential and 
accelerating frames are informationally equivalent. $\bs$

\section{SUPERFLUIDS}
\renewcommand{\theequation}{4.\arabic{equation}}
\setcounter{equation}{0}

The superfluid universe of Klinkhamer and Volovik (cf. 
especially \cite{vovk} (cond-mat 1004.0597) and 
\cite{klvk} (0811.4347) is a fascinating theme and we mention a few 
mathematical points here which were designed to give some structure to the superfluid world.  We have a very imperfect
and fragmentary idea of the physics here and only look at certain
mathematical features which seem related to the material 
sketched above (and in \cite{c006,c007,c009} etc.).
As an example of the arena for ${\mf q}$ theory one looks for
${\mf q}^{\mu\nu}$ which in a homogeneous vacum has the form
$({\bf 4A})\,\,{\mf q}^{\mu\nu}={\mf q}g^{\mu\nu}$ or
$({\bf 4B})\,\,{\mf  q}_{\mu\nu\ga\gb}= 
{\mf q}(g_{\ga\mu}g_{\gb\nu}-
g_{\ga\nu}g_{\gb\mu})$ (generally here $\hbar=c=k_B=1$). 
 Another example is $({\bf 4C})\,\,
{\mf q}^{\mu\nu\ga\gb}={\mf q}\gep^{\mu\nu\ga\gb}$ where
$\gep^{\mu\nu\ga\gb}$ is the fully antisymmetric Levi-Civita tensor (cf. \cite{carm}).  More specifically one can consider a chiral condensate of gauge fields (e.g. a gluonic condensate
in QCD).  Assume e.g. that $F_{\ga\gb}$ represents a color
magnetic field with $({\bf 4D})\,\,<F_{\ga\gb}>=0$ and 
$<F_{\ga\gb}F_{\mu\nu}>=({\mf q}/24)\sqrt{-g}\gep_{\ga\gb\mu\nu}$ where ${\mf q}$ is the anomaly driven topological
condensate 
\bq\label{4.1}
{\mf q}=<\tl{F}^{\mu\nu}F_{\mu\nu}>=\frac{1}{\sqrt{-g}}\gep^{\ga\gb\mu\nu}<F_{\ga\gb}F_{\mu\nu}>
\end{equation}
Then one chooses a vacuum action $({\bf 4E})\,\,S_{{\mf q}}=
\int d^4x\sqrt{-g}\gep({\mf q})$ leading to a stress-energy tensor
\bq\label{4.2}
T^{\mf q}_{\mu\nu}=-\frac{2}{\sqrt{-g}}\frac{\gd S_{\mf q}}{\gd g^{\mu\nu}}=
\gep({\mf {\mf q}})g_{\mu\nu}-2\frac{\pp\gep}{\pp {\mf {\mf q}}}
\frac{\pp {\mf q}}{\pp g^{\mu\nu}}
\Rightarrow
\frac{\pp q}{\pp g^{\mu\nu}}=\frac{1}{2}qg_{\mu\nu}\Rightarrow
\end{equation}
$$\Rightarrow T^{\mf q}_{\mu\nu}=g_{\mu\nu}\rho_{vac}({\mf q});\,\,\rho_{vac}({\mf q})=\gep({\mf q})-{\mf q}
\frac{\pp\gep}{\pp {\mf q}}$$
In terms of the Einstein equations one has then 
$({\bf 4F})\,\,
T^{\mf q}_{\mu\nu}=\gL g_{\mu\nu};\,\,\gL=\rho_{vac}({\mf q})=\gep({\mf q})-{\mf q}\frac{\pp\gep}{\pp {\mf q}}$.  Other possibilities are indicated in \cite{klvk,vovk} but fulfillment of
(4.2) and ({\bf 4F}) is essential.  Recall also from Section 3 that $\gep_{vac}-\mu{\mf q}=-P_{vac}$ where $\mu$ is the chemical
potential and one can argue that 
$-P_{vac}\sim\gL\sim\rho_{vac}$.
\\[3mm]\indent
We recall now, from the discussion in Section 3, that given an osmotic velocity ${\bf u}$
in some region $\gO\in {\bf R}^3$ there will be a local osmotic
pressure 
\bq\label{4.3}
P_{loc}=\frac{1}{2}\rho{\bf u}^2=\frac{m}{2}PD^2\frac{|\na P|^2}
{P^2}=\frac{\hbar^2}{8m}\frac{|\na P|^2}{P}
\end{equation} 
so that $({\bf 4G})\,\,P_{osm}=\int_{\gO}P_ud^3x=\int {\mf Q}Pd^3x=(\hbar^2/8m)FI$.  In the superfluid universe 
mentioned above (with vacuum ${}^4He$) 
one might to suggest that this could correspond to a vacuum osmotic pressure induced by a quantum generated ``vacuum" based on Sections 1-2.  In order to develop this one should
examine the osmotic pressure idea in a 4-D context, in
view of the expression ({\bf 2F}) for ${\mf Q}$.
\\[3mm]\indent
We sketch now from \cite{grsg} (quant-ph 0201035) where
the Nelsonian perspective is extended to a deBroglie-Bohm 
(BB) in the context of describing a KG framework (cf. also \cite{lepk}).
Perhaps somewhat prophetically it is mentioned in \cite{grsg}
(quant-ph 0201035) that the quantum vacuum has to play a
major role in the evolution of the universe and one must take
$\gL$ seriously (which directive is being followed today,
especially in the theory of superfluids).  The introduction of an aether is appropriate in a
relativistic (BB) theory and enters into the KG construction.
We extract now from \cite{grsg} (quant-ph 0201035) a few
equations regarding osmotic velocity in hopes of implementing
in some manner the superfluid theory.  Thus, following \cite
{grsg}, we write $({\bf 4H})\,\,\psi(x,t)=R(x,t)exp(-iS/\hbar);\,\,
[\bx+(m^2c^2/\hbar^2)]\psi=0$ leading to ($\bx=(1/c^2)\pp_t^2
+\pp_x^2+\pp_y^2+\pp_z^2$)
\bq\label{4.4}
\frac{\bx R}{R}+\frac{1}{\hbar}\frac{\pp_{\mu}R}{R}\pp^{\mu}S+
\frac{i}{\hbar}\frac{\pp^{\mu}R}{R}\pp_{\mu}S-\frac{1}{\hbar^2}
\pp_{\mu}S\pp^{\mu}S+\frac{i}{\hbar}\bx S+\frac{m^2c^2}{\hbar^2}=0
\end{equation}
from which follows (for $P=R^2$ and $J_{\mu}=P\pp_{\mu}S$)
\bq\label{4.5}
2\frac{\pp^{\mu}R}{R}\pp_{\mu}S+\bx S=0;\,\,\pp^{\mu}J_{\mu}=\pp^{\mu}P\pp_{\mu}S+P\bx S=0
\end{equation}
along with $({\bf 4I})\,\,\pp_{\mu}S\pp^{\mu}S=M^2c^2=m^2c^2
+\hbar^2(\bx R/R)$.  Now to include vacuum energy one writes
$({\bf 4J})\,\,S^0=E_{vac}t$ along with  
$({\bf 4K})\,\,\Psi(x,t)=R(x,t)
exp[-i(S+S^0)/\hbar]$ for a solution of the KG equation.
Then instead of (4.5) one involves a ``hidden" probability current $J_{\mu}^0=P\pp_{\mu}S^0$ to get a conservation law for the
total probability current $({\bf 4L})\,\,\pp^{\mu}(J_{\mu}+J^0_{\mu})=0$ where $({\bf 4M})\,\,\pp^{\mu}J_{\mu}=\pp^{\mu}P
\pp_{\mu}S+P\bx S=-\pp^{\mu}P\pp_{\mu}S^0-P\bx S^0=-\pp^{\mu}J^0_{\mu}$.  Further $S^0$ also modifies ({\bf 4I}) leading
to a variable total mass 
\bq\label{4.6}
M^2c^2\to\left[mc+\frac{E_{vac}}{c}\right]^2+\hbar^2\frac{\bx R}
{R}
\end{equation}
Thus a (``hidden") diffusion current $J_{\mu}^0$ is introduced,
due to the assumed stochastic aether dynamics and a corresponding all-pervading vacuum energy $E_{vac}$.
One expects that timelike trajectories will ensue, i.e. the 
zero component $J^0_{vac}=P\pp_t(S+S^0)\geq 0$.
\\[3mm]\indent
Now in order to relate this causal
KG theory with Nelson's stochastic theory one defines
$({\bf 4N})\,\,S=Et-{\bf p}{\bf x}=\hbar p_{\mu}x^{\mu}$
such that ${\bf v}^{\mu}=\pp^{\mu}S/M$.  Also an external potential
V is assumed (which could include the vacuum energy $E_{vac}$) with M given by ({\bf 4I}) as $(\bullet)\,\,M=\{[m+(V/c^2)]^2+(\hbar^2/c^2)(\bx R/R)]\}^{1/2}$.  Thus the two real valued
equations corresponding to the KG equation involves $({\bf 4O})\,\,\pp^{\mu}J_{\mu}=\pp^{\mu}P\pp_{\mu}S+P\bx S=0$ and a Hamilton-
Jacobi-Bohm equation
\bq\label{4.7}
\pp_{\mu}S\pp^{\mu}S=M^2{\bf v}_{\mu}{\bf v}^{\mu}=M^2c^2=\left[mc+\frac{V}{c}\right]^2+{\mf Q}
\end{equation}
where $({\bf 4P})\,\,{\mf Q}=\hbar^2(\bx R/R)$.  This leads to
an ``osmotic velocity" $({\bf 4Q})\,\,{\bf u}^{\mu}=
D[\pp^{\mu}P/P]$ with $D=\hbar/2m$ (cf. also \cite{grsg} - book).
One can then rewrite ({\bf 4O}) and (4.7) respectively as
\bq\label{4.8}
\pp_{\mu}{\bf v}^{\mu}=-\frac{1}{D}{\bf u}_{\mu}{\bf v}^{\mu}-\frac{\pp_{\mu}M}{M}{\bf v}^{\mu};\,\,M^2c^2=\left[mc+\frac{V}{c}\right]^2+{\mf Q}
\end{equation}
where $({\bf 4R})\,\,{\mf Q}=m^2c^2+m\hbar\pp_{\mu}{\bf u}^{\mu}$ showing that the dynamics is essentially governed by the
4-gradient of the osmotic velocity.  Also there is then a local
osmotic pressure based on $(1/2)M|{\bf u}|^2$ with M as in
$(\bullet)$.
In the sense of \cite{lepk} (but now in explicit dependence on
the osmotic velocity) equations (4.8) are the two basic equations
of relativistic stochastic mechanics (cf. \cite{grsg} for more on
this).  Note that one can also rewrite (4.8-A) as
\bq\label{4.9}
\pp_{\mu}{\bf v}^{\mu}=-\frac{1}{D}({\bf u}_{\mu}+{\bf u}_{\mu}^M){\bf v}^{\mu};\,\,{\bf u}_{\mu}^M=D\frac{\pp_{\mu}M}{M}
\end{equation}
We refer to \cite{grsg} for more equations and discussion
and anticipate possible interesting connections of the analysis
here to superfluid theory.
\\[3mm]\indent
The obvious question now is to determine a possible relation
between M and ${\mf M}$ where ${\mf M}$ is the quantum mass
of Section 1 and M is the relativistic mass of \cite{grsg}.  We recall that the Weyl geometry involved $\hat{g}_{ab}=\gO^2(x)
g_{ab}$ with$({\bf 4S})\,\,{\mf M}^2=m^2exp({\mf Q}$ and ({\bf 4T}) ${\mf Q}=(\hbar^2/m^2)(\bx |\Psi|/|\Psi|)$.  On the other
hand there is no Weyl geometry explicit in Section 4 and one has
from $(\bullet)$, (4.7), (4.8)
\bq\label{4.10}
M^2c^2=\left[mc+\frac{V}{c}\right]^2+{\mf Q};\,\,{\mf Q}=
\hbar^2\frac{\bx R}{R}
\end{equation}
Now we recall that when working directly with a KG equation
it is natural to use an approximation $({\bf 4U})\,\,exp({\mf Q}\sim
1+{\mf Q}$ (cf. \cite{c006,c007,shoj}) and ({\bf 4S}) becomes then $({\bf 4V})\,\,{\mf M}^2\sim m^2(1+{\mf Q})$.  Setting $c=1$ in (4.10) then one can compare
\bq\label{4.11}
M^2=(m+V)^2+\hbar^2\frac{\bx R}{R}\,\,\,and\,\,\,
{\mf M}^2=m^2\left[1+\frac{\hbar^2}{m^2}\frac{\bx R}{R}\right]
\end{equation}
Evidently ${\mf M}^2\sim M^2$ with $m\rightarrow m+V$
where V contains $E_{vac}$.  If we take $V=E_{vac}$ then M
only differs from ${\mf M}$ by an $E_{vac}$ term, i.e. for 
$\tl{{\mf Q}}=\hbar^2\bx R/R$
\bq\label{4.12}
\sqrt{M^2-\tl{{\mf Q}}}-\sqrt{{\mf M}^2-\tl{{\mf Q}}}=E_{vac}
\end{equation}
In a certain sense then the effect of relativity here is to
display the influence of a quantum vacuum.


\newpage
\begin{thebibliography}{cccc}

%
\bibitem{adsr} R. Adler, M. Bazin, and M. Schiffer,
Introduction to general relativity, McGraw Hill, 1965
%
\bibitem{audr} J. Audretsch, Phys. Rev. D, 24 (1981), 1470-1477;
27 (1983), 2872-2884
%
\bibitem{agsn} J. Audretsch, F. G\"aehler, and N. Straumann, Comm. Math. Phys.,
95 (1984), 41-51
%
\bibitem{aula} J. Audretsch and C. L\"ammerzahl, Class. Quant. Gravity, 5 (1988),
1285-1295
%
\bibitem{bdmh} C. Bender and P. Mannheim, Phys. Rev. D, 78 (2008), 025022;
Phys. Rev. Lett., 100 (2008), 110402; hep-th 0807.2607
%
\bibitem{brfm} G. Bertoldi, A. Faraggi, and M. Matone,
Class. Quant. Grav., 17 (2000), 3965
%
\bibitem{bmhl} D. Bohm and B. Hiley, Phys. Repts., 172 (1989), 93-122
%
\bibitem{bqcd} R. Bonal, I. Quiros, and R. Cardenas, gr-qc 0010010
%
\bibitem{cai} R. Cai and S. Kim, hep-th 05001055
%
\bibitem{caoh} R. Cai, L. Cao, and N. Ohta, hep-th 1001.3470 and 1002.1136
%
\bibitem{chla} S. deCampo, R. Herrera, and P. Lebrana, gr-qc 0711.1559
%
\bibitem{caht} V. Canuto, P. Adams, S. Hsieh, and E. Tsiang,
Phys. Rev. D, 16 (1977), 1643-1663
%
\bibitem{ccsh} Q. Cao, Y. Chen, and K. Shao, hep-th 1001.2597
%
\bibitem{carm} M. Carmelli, Classical fields, World Scientific,
2001
%
\bibitem{c006} R. Carroll, Fluctuations, information, gravity, and the quantum
potential, Springer, 2006
%
\bibitem{c007} R. Carroll, On the quantum potential, Arima
Publ., 2007
%
\bibitem{c009} R. Carroll, On the emergence theme of physics,
World Scientific, 2010
%
\bibitem{caar} R. Carroll, Progress in Physics, 4 (2007), 22-24;
1 (2008), 21-24 and 2 (2008), 89-90
%
\bibitem{caro} R. Carroll, math-ph 0701077. 0703065, 0705.3921, 0712.3251; 0507027
%
\bibitem{cl} R. Carroll, Foundations of Physics, 35 (2005), 131-154
%
\bibitem{crcl} R. Carroll, Teor. i Mat. Fiz., 152 (2007), 904-914
%
\bibitem{crrl} R. Carroll, math-ph 0807.1320 and 0807.4158
%
\bibitem{crar} R. Carroll, Quantum Potential as Information: A mathematical survey,  in:
New trends in quantum information, Eds. Felloni, Singh, Licata, and Sakaji, Aracne Editrice,
2010, pp. 155-189
\bibitem{clrr} R. Carroll, math-ph 1007.4744; gr-qc 1010.1732
%
\bibitem{cstr} C. Castro, Found. Phys., 22 (1992), 569-615 and
37 (2007), 366-409;
Found. Phys. Lett., 4 (1991), 81-99; Jour. Math. Phys., 31 (1990), 2626-2633
%
\bibitem{cath} A. Caticha, quant-ph 0907.4335, 1005.2357, and 1011.0746
%
\bibitem{crow} L. Crowell, Quantum fluctuations of spacetime, World Scientific, 2005
%
\bibitem{dirc} P. Dirac, Proc. Royal Soc. London, A 333 (1973),
403 and 338 (1974), 439
%
\bibitem{drlr} W. Drechsler, gr-qc 9901030
%
\bibitem{ezaw} Z. Ezawa, Quantum Hall effects, World Scientific,
2008
%
\bibitem{fons} F. Falciano, M. Novello, and J. Salim, Found. Phys.,
40 (2010), 1885-1901
%
\bibitem{fgme} A. Faraggi and M. Matone, Inter. Jour. Mod. Phys.
A, 15 (2000), 1869-2017
%
\bibitem{faro} V. Faraoni, Cosmology in scalar-tensor theory,
Kluwer, 2004; gr-qc 0906.1901 and 1005.2327
%
\bibitem{fred} B. Frieden, Science from Fisher information, Springer, 2004; Physics
from Fisher information, Cambridge Univ. Press, 1998
%
\bibitem{fpps} B. Frieden, A. Plastino, A. R. Plastino, and B. Soffer,
Phys. Rev. E, 60 (1999), 48-53 and 66 (2002), 046128; Jour. Phys. A, 304 (2002), 73-78
%
\bibitem{frgy} B. Frieden and R. Gatenby, Exploratory data analysis using Fisher information,
Springer, 2007
%
\bibitem{fnsf} B. Frieden and B. Soffer, Physica A, 388 (2009), 1315-1330
%
\bibitem{fumd} Y. Fuji and K. Maeda, The scalar-tensor theory of gravitation, Cambridge Univ. Press, 2003
%
\bibitem{garb} P. Garbaczewski, Entropy, 7 (2005), 253-299; Jour. Stat. Phys. 123 (2007),
315-355; quant-ph 0408192, 0612151, 0706.2481, and 0805.1536; cond-mat 0211362, 0604538, 0703147, 0703204, and 0811.3856
%
\bibitem{glkn} L. Glinka, gr-qc 0612079, 0707.3341, 0711.1380, 0712.1624, 0712.2769,
0801.4157, 0803.0533, and 0808.1533; physics 1102.5002 (book)
%
\bibitem{ggbv} M. Gogberashvili, physics 0910.0169; gr-qc 
0807.2439 and1008.2544
%
\bibitem{gbkv} M. Gogberashvili and I. Kanatchikov, gr-qc
0807.2439 and 1009.2266; physics 1012.5914
%
\bibitem{graf} W. Graf, gr-qc 0602054
%
\bibitem{grne} B. Greene, The hidden reality, Knopf, 2011
%
\bibitem{gros}
G. Gr\"ossing, Entropy, 12 (2010), 1975-2044;  quant-ph 
0201035, 0205048, 0311109, 0404030, 0410236, 0508079, 0711.4954, 0806.4462, and 0808.3539;  Phys. Lett. A, 372 (2008), 4556-4562; Found.
Phys. Lett., 17 (2004), 343-362
%
\bibitem{grsg} G. Gr\"ossing, Quantum cybernetics, Springer,
2000; quant-ph 0201035 and 0205047
%
\bibitem{gfps} G. Gr\"ossing, S. Fussy, J. Mesa Pascasio, and H. Schwabl,
quant-ph 1004.4596 and 1005.1058
%
\bibitem{gsps} G. Gro\"ssing and J. Mesa Pascasio, quant-ph 0812.3561
%
\bibitem{hall} M. Hall, quant-ph 9802052, 9806013, 9903045, 9912055, 0103072, 0107149, 0302007;
gr-qc 0408098; Phys. Rev. A, 55 (1997), 100-113 and 62 (2000), 012107, Jour. Phys. A, 37 (2004), 7799;
Gen. Relativ. Grav., 37 (2005), 1505-1515
%
\bibitem{hlrg} M. Hall and M. Reginatto, Jour. Phys. A. 35 (2002), 3289-3303; Fortschrift
Phys., 50 (2002), 646-651
%
\bibitem{hkrg} M. Hall, K. Kumar, and M. Reginatto, Jour. Phys. A, 36 (2003), 9779-9794
(hep-th 0206235 and 0307259)
%
\bibitem{hark} T. Harko, gr-qc 1101.3655
%
\bibitem{isfc} J. Isidro, J. Santander, and P. Fernandez de
Cordoba, hep-th 0902.0143 and 0912.1535
%
\bibitem{isrt} M. Israelit, The Weyl-Dirac theory and our
universe, Nova Sci. Publ., 1999
%
\bibitem{istr} M. Israelit, Found. Phys., 29 (1999), 1303; 32 (2002), 295 and 945; 28 (1998), 205; 35 (2005), 1725 and 1769; 37 (2007), 1628; gr-qc 9609035, 0710.3709, 0710.3910, 07103913, 07103923, and
0905.2482; astro-ph 1008.0767; Inter. Jour. Mod. Phys. A, 17
(2002), 4229
%
\bibitem{isrn} M. Israelit and N. Rosen, Found. Phys., 13 (1983), 1023
%
\bibitem{jacb} T. Jacobson, Phys. Rev. Lett., 75 (1995), 1260;
hep-th 1001.4853; gr-qc 0711.3852 and 0801.1547
%
\bibitem{joct} D. Johnson and A. Caticha, quant-ph 1010.1467
%
\bibitem{khol} A. Kholodenko, gr-qc 0110064; hep-th
0303334 and 0701084
%
\bibitem{klle} H. Kim, J. W. Lee, and J. Lee, 0709.3573
%
\bibitem{klhr} F. Klinkhamer, hep-th 1001.1939, 1006.2094,
1101.1281, and 1101.5370; gr-qc 0703009
%
\bibitem{klvk} F. Klinkhamer and G. Volovik, gr-qc 0711.3170, 0807.3896, 0811.4347, and 1102.3152; hep-th 0905.1919 and 0907.4887
%
\bibitem{leew} J.W.  Lee, hep-th 1003.1878, 1003.4464,1005.2739, and 1011.1657
%
\bibitem{lkle} J.W. Lee, H. Kim and J. Lee, hep-th 0701199, 1001.5445 and 1002.4568
%
\bibitem{lepk} W. Lehr and J. Park, Jour. Math. Phys., 18 (1977),
1235-1240
%
\bibitem{ldhd} B. Liu, Y. Dai, X. Hu, and J. Deng, gr-qc 1012.2560
%
\bibitem{makl} J. Makela, gr-qc 0506087, 0605098,  061078, 0805.3952, 0805.3955, 0810.4901, 1001.3808, and 1011.2052
%
\bibitem{mgsk} G. Mangano and L. Sokolowski, gr-qc 9312008
%
\bibitem{mnhm} P. Mannheim, hep-th 0707.2283, 0809.1200, 0909.0212, 
0912.2635, 1005.5108, and 1101.2186; astro-ph 0505266
gr-qc 0001011 and 0703037
%
\bibitem{mkmr} J. Munkhammar, hep-th 1003.1262 and 1006.0723; physics 1003.4981 and 1101.1417
%
\bibitem{naga} M. Nagasawa, Schroedinger equations and diffusion theory, 
Birkh\"auser, 1993; Stochastic processes in quantum physics, Birkh\"auser, 2000
%
\bibitem{nels} E. Nelson, Quantum fluctuations, Princeton Univ. Press, 1985; Dynamical
theory of Brownian motion, PUP, 1967; Phys. Rev., 130 (1966), 1079-1085
%
\bibitem{novl} M. Novello, astro-ph.CO, 1003.5126; physics 1004.3195 and 1008.2371
%
\bibitem{nsfo} M. Novello, J. Salim, and F. Falciano, gr-qc 0901.3741; Found. Phys., 40 (2010), 1885-1901
%
\bibitem{pada} T. Padmanabhan, Gravitation, Cambridge Univ. Press, 2010
%
\bibitem{padn} T. Padmanabhan, gr-qc 0202078, 0209088, 0308070, 0311036, 0408051, 0412068, 0606061, 0910.0839, 0911.1403, 0911.5004, 0912.3165, 1001.3380, 1003.5662,  1007.5066, 1012.0119, and 1012.4476; hep-th 0205278
%
\bibitem{papn} G. Papini, gr-qc 0111005 and 0304082
%
\bibitem{ppwd} G. Papini and W. Wood, Phys. Lett. A, 202
(1995), 45-49
%
\bibitem{ptsh} C. Pethick and H. Smith, Bose Einstein condensation in dilute gases, Cambridge Univ. Press, 2008
%
\bibitem{pist} L. Pitaevskij and S. Stringari, Bose-Einstein 
condensates, Oxford Univ. Press, 2003
%
\bibitem{quir} I. Quiros, gr-qc 9904004, 9905071, and 0004014; hep-th 0009169
%
\bibitem{regn} M. Reginatto, Phys. Rev. A, 58 (1998), 1775; quant-ph 9909065;
gr-qc 0501030
%
\bibitem{relg} M. Reginatto and F. Lengyel, cond-mat 9910039
%
\bibitem{rosn} N. Rosen, Found. Phys., 12 (1982), 213
%
\bibitem{sant} E. Santamato, Phys. Rev. D, 29 (1984), 216-222 and
32 (1985), 2615-2621; Jour. Math. Phys., 25 (1984), 2472-2480;
Phys. Lett. A, 130 (1988), 199-202
%
\bibitem{sclz} E. Scholz, astro-ph 0403446 and 0805.2557; hep-th 1102.3478; gr-qc 0703102 and 0710.0269
%
\bibitem{shoj} F. and A. Shojai, gr-qc 0306099, 0404102, 0409053 and 001.3496; Inter. Jour. Mod. Phys. A, 15 (2000), 1859; Class. ,Quant. Grav., 21 (2004), 1-9
%
\bibitem{smol} L. Smolin, hep-th 0605052; quant-ph 0609109; gr-qc 1001.3668; Class. Quant. Grav., 3 (1986), 347-359; Phys. Lett., 113A (1986), 408-412
%
\bibitem{vlde} E. Verlinde, hep-th 1001.0785
%
\bibitem{volv} G. Volovik, The universe in a helium droplet,
Oxford Univ. Press, 2003
%
\bibitem{vovk} G. Volovik, gr-qc 0004049, 0405012, 0406005, 0505104,
and 0612134;
cond-mat 0507454 and 1004.0597; physics 0909.1044;
Found. Phys., 41 (2011), 516-528
%
\bibitem{wanc} C. Wang, gr-qc 0603062 and 0605124
%
\bibitem{wlwg} S. Wei, Y. Liu, and Y. Wang, hep-th 1001.5238
%
\bibitem{wein} S. Weinberg, Cosmology, Oxford Univ. Press,
2008
%
\bibitem{whlr} J. Wheeler, Phys. Rev. D, 41 (1990), 431-441 and
44 (1991), 1769-1773
%
\bibitem{wdpp} W. Wood and G. Papini, Phys. Rev. D, 45
(1992), 3617-3627; Found. Phys. Lett., 6 (1993), 207-223;
Phys. Lett. A, 202 (1995), 45-49; gr-qc 9612042
%
\bibitem{xiog} H. Xiong, hep-th 1012.5858, 1101.0525, and 1101.1270
%








\end{thebibliography}
\end{document}